\documentclass[12pt]{article}
\usepackage{epsfig}
\usepackage{amsfonts}
\usepackage{latexsym}
\usepackage{amsmath}
\usepackage{mathrsfs}
\usepackage{hyperref}
\usepackage{setspace}
\usepackage{color}
\usepackage{bm}
\usepackage{slashed}
\numberwithin{equation}{section}

\begin{document}

\begin{titlepage}
\unitlength = 1mm
\begin{flushright}
YITP-16-136\\
\end{flushright}

\vskip 1cm
\begin{center}

{\Large {\textsc{\textbf{Vacuum State of the Dirac Field in de Sitter Space\\
 and Entanglement Entropy}}}}

\vspace{1.8cm}
Sugumi Kanno$^{*\,\flat}$, Misao Sasaki$^{\natural}$ and
 Takahiro Tanaka$^{\dag\,\natural}$

\vspace{1cm}

\shortstack[l]
{\it $^*$ Department of Theoretical Physics and History of Science,
University of the Basque Country\\
~~48080 Bilbao, Spain\\
\it $^\flat$ IKERBASQUE, Basque Foundation for Science, 
Maria Diaz de Haro 3,
48013, Bilbao, Spain\\
\it $^\natural$ Center for Gravitational Physics,\\
~~Yukawa Institute for Theoretical Physics, Kyoto University,
Kyoto 606-8502, Japan\\
\it $^\dag$ Department of Physics, Kyoto University,
Kyoto 606-8502, Japan
}

\vskip 1.5cm

\begin{abstract}
\baselineskip=6mm
We compute the entanglement entropy of a free massive Dirac field 
between two causally disconnected open charts in de Sitter space. 
We first derive the Bunch-Davies vacuum mode functions of the Dirac field. 
We find there exists no supercurvature mode for the Dirac field.
We then give the Bogoliubov transformation between the Bunch-Davies 
vacuum and the open chart vacua that makes the reduced density 
matrix diagonal. We find that the Dirac field becomes more entangled than 
a scalar field as $m^2/H^2$ becomes small, and the difference is maximal
in the massless limit.
\end{abstract}

\vspace{1.0cm}

\end{center}
\end{titlepage}

\pagestyle{plain}
\setcounter{page}{1}
\newcounter{bean}
\baselineskip18pt

\setcounter{tocdepth}{2}

\tableofcontents

\section{Introduction}

Since when Einstein-Podolsky-Rosen (EPR) pointed out in 1935, 
quantum entanglement has fascinated many physicists 
because of its counterintuitive nature that a local measurement on 
a particle may affect the outcome of a local measurement
on a distant particle instantaneously~\cite{Einstein:1935rr}.
 After Aspect et al. convincingly tested the quantum nature
of entanglement by measuring correlations 
of linear polarizations of pairs of photons~\cite{Aspect:1981zz, Aspect:1982fx}, 
much attention has been paid to this genuin quantum property 
in various research areas including
quantum information theory, quantum communication, 
quantum cryptography, quantum teleportation 
and quantum computation. 
 
Turning our eyes on cosmology, in de Sitter space where
the universe expands exponentially, any two of mutually separated
region eventually becomes causally disconnected. This is most
conveniently described by spanning open universe coordinates
on two open charts in de Sitter space. 
The positive frequency mode functions of a free massive scalar field 
for the Euclidean vacuum (the Bunch-Davies vacuum) that have support 
on both regions were derived in~\cite{Sasaki:1994yt}. 
Using them, several studies have been made on the quantum 
entanglement, particularly, entanglement entropy~\cite{Maldacena:2012xp}, 
on negativity~\cite{Kanno:2014bma}, which is a measure of entanglement for any mixed states involved 
in subsystems, and on quantum discord~\cite{Kanno:2016gas}.

Quantum entanglement between two causally
 disconnected regions in de Sitter space was first studied
by Maldacena and Pimentel~\cite{Maldacena:2012xp}. 
They showed that the entanglement entropy,
 which is a measure of quantum entanglement, of a free massive 
scalar field between two disconnected open charts is non-vanishing. Motivated by
this result, there have been several attempts to test the idea of multiverse by 
studying long range correlations of various states that quantum entanglement
naturally gives rise to~\cite{Kanno:2014ifa, Kanno:2015lja, Kanno:2015ewa}. 

In order to gain some insight into relativistic quantum information,
qunatum entanglement between causally disconnected regions in flat space
was investigated by Fuentes-Schuller and Mann~\cite{FuentesSchuller:2004xp}
by making use of the Rindler coordinates.
They studied the entanglement between two causally disconnected modes
 of a free scalar field as viewed by two relatively accelerated observers 
in Rindler space. It was found that the entanglement is degraded from 
the perspective of accelerated observers in flat space and, in particular, 
the entanglement disappears for an infinitely accelerated observer.
Interestingly, however, Alsing et al. showed that unlike bosonic fields
fermionic fields always remain entangled even in the limit of infinite 
acceleration~\cite{Alsing:2006cj}. Then Datta showed that
quantum discord, which is a measure of all quantum correlations 
including quantum entanglement, never disappears in this limit~\cite{Datta}.

The two open charts of de Sitter space are analogous to the Rindler wedges 
in flat space in the sense that an observer
in the region described by one of the charts has no access to the field modes 
in the other causally disconnected region. Therefore it is of interest
to see how the spacetime curvature will affect these results obtained in flat
space.

In this paper, we first study the Bunch-Davies vacuum of
the Dirac field in open charts by extending the previous work on scalar fields
~\cite{Sasaki:1994yt}. Then using thus obtained spinor mode functions, 
we compute the entanglement entropy.

The paper is organized as follows. In section 2, we consider the 
Dirac equation and its conserved currents in the open chart. In section 3, 
we derive the mode functions in each of the two open chart regions. 
In section 4, we analytically extend the solutions to construct
the positive frequency mode functions for the Bunch-Davies vacuum. 
In section 5, we present the Bunch-Davies vacuum solutions in the open chart. 
In section 6, we discuss the absence of supercurvature modes for fermions. 
In section 7, we compute the entanglement entropy between the two
causally separated open charts.
Finally we summarize our reslut and discuss the implications in section 8.

\section{The Dirac equation in open chart}
\begin{figure}[t]
\vspace{-3cm}
\includegraphics[height=11cm]{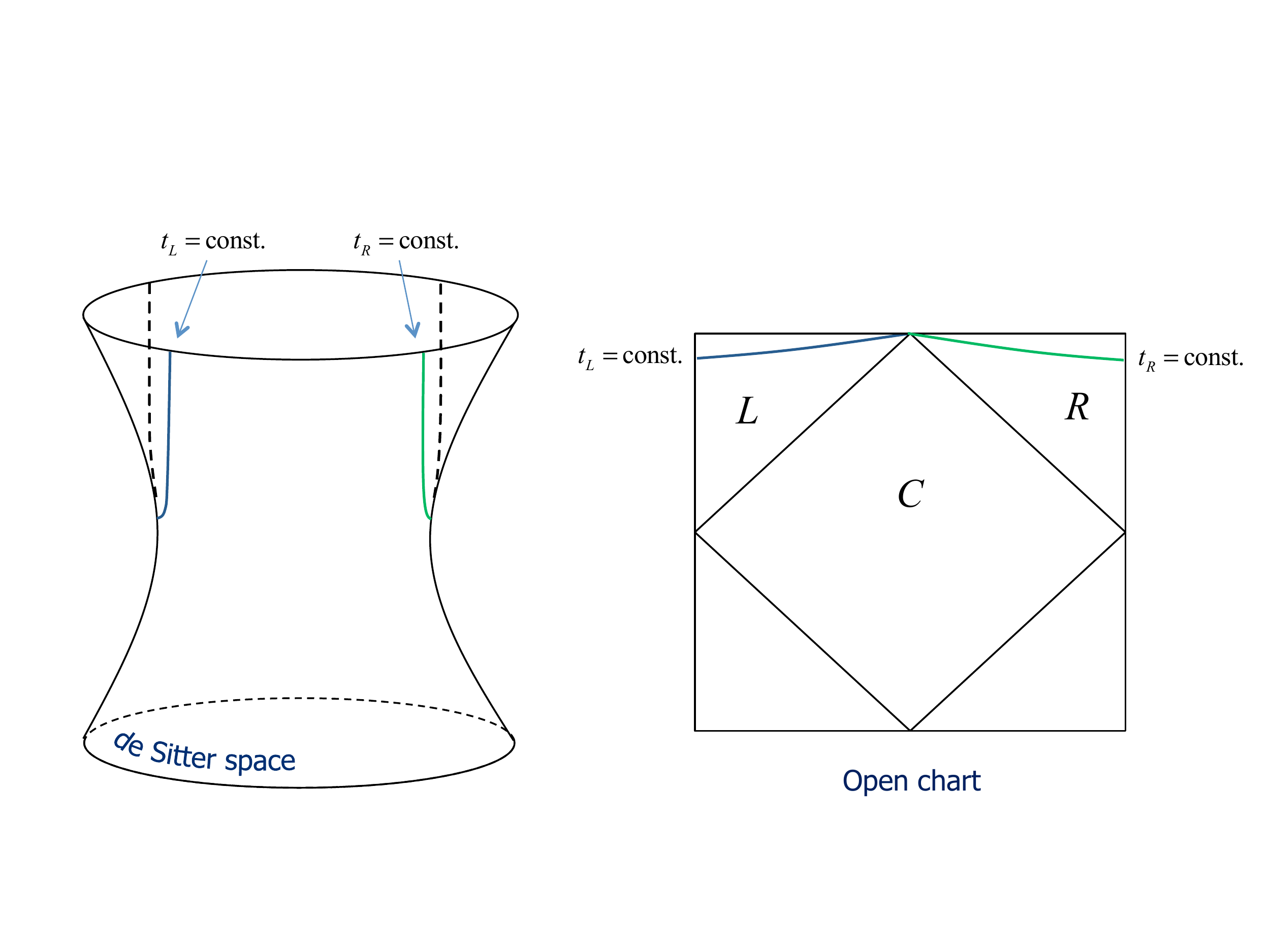}\centering
\vspace{-1cm}
\caption{De Sitter space and its Penrose diagram. 
The regions $R$ and $L$ are the two causally disconnected open 
charts of de Sitter space.}
\label{fig1}
\end{figure}

The open de Sitter space in the case of a free massive scalar field 
is studied in detail in~\cite{Sasaki:1994yt}. In Figure~\ref{fig1}, 
the de Sitter space and its Penrose diagram are depicted. 
We extend the study~\cite{Sasaki:1994yt} to the case of a Dirac 
field in this section. We do not include dynamical gravity 
as in~\cite{Sasaki:1994yt}.

The metric in the open chart is
\begin{eqnarray}
ds_4^2&=&g_{\mu\nu}\,dx^\mu dx^\nu
=H^{-2}\left[-dt^2+\sinh^2t\,dS_3^2\right]\,,\nonumber\\
&\equiv&\eta_{AB}\,e^A\,e^B
=-\left(e^0\right)^2+\sinh^2t\,\delta_{ab}\,{\tilde e}^a\,{\tilde e}^b\,,
\label{metrics}
\end{eqnarray}
where the indices $(\mu\,,\nu)$ run from 1 to 4 and $H^{-1}$ 
is the curvature radius of the de Sitter space. The second line is 
the metric in the tetrad system and the indices $(A,B)$ and $(a,b)$ 
run from 1 to 4 and 1 to 3, respectively. The relation between 
curved and flat metrics is given by
 $g_{\mu\nu}=\eta_{AB}\,e_{\mu}{}^{A}\,e_{\nu}{}^{B}$ where 
the tetrad field satisfies $e^A=e_{\mu}{}^{A}dx^\mu$. 
We defined $e^a=\sinh t\,{\tilde e}^a$.

We choose $4\times4$ gamma matrices in accordance with our 
metric signature $(-,+,+,+)$ such that
\begin{eqnarray}
\gamma^0=\left(
\begin{array}{cc}
0 & iI\\
iI & 0
\end{array}
\right)\,,\qquad
\gamma^a=\left(
\begin{array}{cc}
0 & i\tilde\gamma^a\\
-i\tilde\gamma^a & 0
\end{array}
\right)\,,
\end{eqnarray}
where $I$ and $\tilde\gamma^a$ are the $2\times 2$ unit matrix and gamma matrices respectively. 
The gamma matrices satisfy $\left\{\gamma^A\,,\gamma^B\right\}=2\eta^{AB}$.

The Dirac equation for a four dimensional massive spinor $\Psi$ is
 then expressed as $\left[\gamma^\mu D_\mu-m\right]\Psi=0$ where
 covariant derivative is
 $D_\mu=\partial_\mu+\frac{1}{2}\omega_{\mu EF}\Sigma^{EF}$ 
involving the spin connection $\omega_{\mu EF}$, 
\begin{eqnarray}
\left[\gamma^A\left(e_{A}{}^\mu\partial_\mu
+\frac{1}{2}\gamma_{EFA}\Sigma^{EF}\right)-m\,\right]\Psi=0\,.
\end{eqnarray}
Here commutators are $\Sigma^{AB}=\frac{1}{4}\left[\gamma^A,~\gamma^B\right]$ 
and the spin connection is given by 
$\omega^{A}{}_B=\gamma^A{}_{BC}\,e^{C}=\gamma^A{}_{B\mu}dx^\mu$. 
If we use Eq.~(\ref{metrics}), the above equation becomes
\begin{eqnarray}
H\gamma^0\left(\partial_t+\frac{3}{2}\frac{f^\prime}{f}\right)\Psi+\frac{H}{f}
\left(
\begin{array}{cc}
0 & i\tilde{\slashed\nabla}\\
-i\tilde{\slashed\nabla} & 0
\end{array}
\right)\Psi-m\Psi=0\,,
\label{dirac}
\end{eqnarray}
where a prime denotes the derivative with respect to the time $t$, 
$f(t)=\sinh t$, $m$ is the mass of $\Psi$ and
 $\tilde{\slashed\nabla}=\tilde{\gamma}^a\tilde{\nabla}_a$.

If we define\footnote{The indices 
$p,\ell,m$ of $\phi_{\pm p\ell m} (t,\Omega)$ are omitted for
 simplicity unless there may be any confusion.}
\begin{eqnarray}
\Psi\left(t,\Omega\right)&=&
\left(
\begin{array}{l}
\phi_{+}\left(t,\Omega\right)\,\\
\phi_{-}\left(t,\Omega\right)
\end{array}\right)\,,
\label{phipm}
\end{eqnarray}
where $\Omega$ is the three-dimensional angle, then the Dirac equation becomes
\begin{eqnarray}
&&iH\left(\partial_t+\frac{3}{2}\frac{f'}{f}\right)\phi_{-}
+i\frac{H}{f}\tilde{\slashed\nabla}\phi_{-}-m\,\phi_{+}=0\,,
\label{dirac1}\\
&&iH\left(\partial_t+\frac{3}{2}\frac{f'}{f}\right)\phi_{+}
-i\frac{H}{f}\tilde{\slashed\nabla}\phi_{+}-m\,\phi_{-}=0\,.
\label{dirac2}
\end{eqnarray}
Combining Eqs.~(\ref{dirac1}) with (\ref{dirac2}), the equation for 
$\phi_+$ is given by
\begin{eqnarray}
\left[\left(\partial_t+\frac{3}{2}\frac{\cosh t}{\sinh t}\right)^2
+\frac{\cosh t}{\sinh^2t}\,\tilde{\slashed\nabla}
-\frac{1}{\sinh^2t}\,\tilde{\slashed\nabla}^2
+\frac{m^2}{H^2}\,\right]\phi_+\left(t,\Omega\right)=0\,.
\label{plus}
\end{eqnarray}
Once we obtain a solution for $\phi_+$, we can get a solution for 
$\phi_-$ by using Eq.~(\ref{dirac2}).

The Dirac equation gives its conserved Noether current expressed as
\begin{eqnarray}
\partial_\mu\left[\sqrt{-g}\,J^\mu\,\right]=0\,,\qquad
J^\mu=-\bar\Psi\gamma^\mu\Psi\,.
\end{eqnarray}
Here, the Dirac adjoint is defined by $\bar\Psi\equiv\Psi^\dag\gamma^0$. 
Then the charge density $(\mu=0)$ is given by
\begin{eqnarray}
J^0=\int dtd\Omega\sqrt{-g}\,\Psi^\dag(t,\Omega)\Psi(t,\Omega)\,,
\end{eqnarray}
where we used $(\gamma^0)^2=-1$.
We use the charge density for mode functions to have orthonormality 
relations below.

\section{The mode function in each $R$ or $L$ region}

In this section, we consider the mode functions in the region $R$ or $L$. 
Since these two regions are completely symmetric, the argument in this 
section can be applied to both $R$ and $L$ regions although we don't 
specify the region $R$ or $L$ below.

\subsection{Positive frequency mode}

We can separate variables according to~\cite{Camporesi:1995fb}:
\begin{eqnarray}
\left(
\begin{array}{l}
\phi_{+}\left(t,\Omega\right)\,\\
\phi_{-}\left(t,\Omega\right)
\end{array}\right)
=\left(
\begin{array}{l}
\phi_{p}(z)\,\chi^{(-)}_{p\ell m}(\Omega)\,\\
\varphi_{p}(z)\,\chi^{(-)}_{p\ell m}(\Omega)
\end{array}\right)\,,
\label{separation1}
\end{eqnarray}
where the three-dimensional spinors $\chi^{(-)}_{p\ell m}(\Omega)$ have
 two components and satisfy
\begin{eqnarray}
\tilde{\slashed\nabla}\chi^{(-)}_{p\ell m}\left(\Omega\right)
=-ip\,\chi^{(-)}_{p\ell m}\left(\Omega\right)\,,
\label{chi1}
\end{eqnarray}
where $p$ is positive and continuous. 
The spinors $\chi^{(-)}_{p\ell m}(\Omega)$ are normalized as
\begin{eqnarray}
\int d\Omega\,\chi_{p\ell m}^{(-)\dag}\,\chi^{(-)}_{p'\ell' m'}
=\delta(p-p')\,\delta_{\ell\ell'}\,\delta_{mm'}\,.
\label{chi:normalization}
\end{eqnarray}
Plugging Eq.~(\ref{separation1}) into Eq.~(\ref{plus}), we get
\begin{eqnarray}
\chi^{(-)}_{p\ell m}\left(\Omega\right)
\left[\left(\partial_t+\frac{3}{2}\frac{\cosh t}{\sinh t}\right)^2
-ip\,\frac{\cosh t}{\sinh^2t}
+\frac{p^2}{\sinh^2t}
+\frac{m^2}{H^2}\,\right]\phi_p(t)=0\,.
\label{plus1}
\end{eqnarray}
Once we obtain a solution for $\phi_p(t)$, we can get a solution 
for $\phi_-\left(t,\Omega\right)=\varphi_p(t)\chi^{(-)}_{p\ell m}(\Omega)$ 
by using Eq.~(\ref{dirac2}).

The positive frequency solution $\Psi^{+}$ is found to be
\begin{eqnarray}
\Psi^{+}_{\uparrow}\left(z,\Omega\right)&=&
\left(
\begin{array}{l}
\phi_{p}(z)\,\chi^{(-)}_{p\ell m}(\Omega)\,\\
\varphi_{p}(z)\,\chi^{(-)}_{p\ell m}(\Omega)
\end{array}\right)
\label{up1}
\end{eqnarray}
where we defined $z\equiv\cosh t$ and the subscript $\uparrow$ indicates spin-up,
\begin{eqnarray}
\phi_p(z)&=&\left(z^2-1\right)^{-\frac{3}{4}}\left(\frac{z+1}{z-1}\right)^{i\frac{p}{2}}
F\left(-i\frac{m}{H}\,,\,i\frac{m}{H}\,,\,\frac{1}{2}-ip\,,\,\frac{1-z}{2}\right)
\label{phi}\,,\\
\varphi_p(z)&=&-\frac{1}{2}\frac{i\frac{m}{H}}{\frac{1}{2}-ip}\left(z^2-1\right)^{-\frac{1}{4}}\left(\frac{z+1}{z-1}\right)^{i\frac{p}{2}}F\left(1-i\frac{m}{H}\,,\,1+i\frac{m}{H}\,,\,\frac{3}{2}-ip\,,\,\frac{1-z}{2}\right)\,.
\label{varphi}
\end{eqnarray}
where $p>0$. Note that $\phi_p$ and $\varphi_p$ realize positive frequency
 in the distant past $t\rightarrow 0~ (z\rightarrow 1)$.

This solution is normalized such that
\begin{eqnarray}
\left(\Psi_{\uparrow}^+,\Psi_{\uparrow}^+\right)
=\delta(p-p')\,\delta_{\ell\ell'}\,\delta_{mm'}\,
\label{norm}
\end{eqnarray}
where the Dirac inner product for the mode function is given by
\begin{eqnarray}
\left(\Psi_\uparrow,\Psi_\uparrow\right)
=\int dtd\Omega\sqrt{-g}\,\Psi^{+\dag}\Psi^+\,,
\end{eqnarray}
at $z\rightarrow 1$.

\subsection{Negative frequency mode}

The negative frequency mode function of Eq.~(\ref{plus}) is given by 
separating variables as in~\cite{Camporesi:1995fb} 
\begin{eqnarray}
\left(
\begin{array}{l}
\phi_{+}\left(t,\Omega\right)\,\\
\phi_{-}\left(t,\Omega\right)
\end{array}\right)
=\left(
\begin{array}{r}
\phi^*_{p}(z)\,\chi^{(+)}_{p\ell m}(\Omega)\,\\
-\varphi^*_{p}(z)\,\chi^{(+)}_{p\ell m}(\Omega)
\end{array}\right)
\equiv\Psi^-_\uparrow\left(z,\Omega\right)\,,
\label{up2}
\end{eqnarray}
where the three-dimensional spinors $\chi^{(+)}_{p\ell m}(\Omega)$ 
have two components and satisfy
\begin{eqnarray}
\tilde{\slashed\nabla}\chi^{(+)}_{p\ell m}\left(\Omega\right)
=ip\,\chi^{(+)}_{p\ell m}\left(\Omega\right)\,,
\label{chi2}
\end{eqnarray}
where $p$ is positive and continuous and the spinors 
$\chi^{(+)}_{p\ell m}(\Omega)$ are normalized in the same way as 
Eq.~(\ref{chi:normalization}). Note that the spinors $\chi^{(+)}_{p\ell m}$ 
are obtained by changing the sign $p\rightarrow -p$ of $\chi^{(-)}_{p\ell m}$. 

Eq.~(\ref{plus1}) corresponding to the negative frequency mode is
\begin{eqnarray}
\chi^{(+)}_{p\ell m}\left(\Omega\right)
\left[\left(\partial_t+\frac{3}{2}\frac{\cosh t}{\sinh t}\right)^2
+ip\,\frac{\cosh t}{\sinh^2t}
+\frac{p^2}{\sinh^2t}
+\frac{m^2}{H^2}\,\right]\phi^*_p(t)=0\,.
\label{plus2}
\end{eqnarray}
Thus, the negative frequency mode function is obtained by replacing
 Eqs.~(\ref{phi}) and (\ref{varphi}) by $p\rightarrow -p$. We find
\begin{eqnarray}
\phi^*_{p}(z)&=&
\left(z^2-1\right)^{-\frac{3}{4}}\left(\frac{z+1}{z-1}\right)^{-i\frac{p}{2}}
F\left(-i\frac{m}{H}\,,\,i\frac{m}{H}\,,\,\frac{1}{2}+ip\,,\,\frac{1-z}{2}\right)
\label{psi}\\
\varphi^*_{p}(z)&=&
\frac{1}{2}\frac{i\frac{m}{H}}{\frac{1}{2}+ip}\left(z^2-1\right)^{-\frac{1}{4}}
\left(\frac{z+1}{z-1}\right)^{-i\frac{p}{2}}F\left(1-i\frac{m}{H}\,,
\,1+i\frac{m}{H}\,,\,\frac{3}{2}+ip\,,\,\frac{1-z}{2}\right)
\label{upsilon}
\end{eqnarray}
where $p>0$ and $*$ denotes complex conjugation. Here 
$\phi^*_{p}$ and $-\varphi^*_p$ realize negative frequency at 
$t\rightarrow 0~ (z\rightarrow 1)$. This solution is normalized 
in the same way as Eq.~(\ref{norm}).

Note that the negative frequency mode is not simply the complex 
conjugate of the positive frequency mode but a minus sign is necessary
 for the lower 2 components.

\subsection{Spin-down solutions for the positive and negative mode functions}
If we take negative $p$ and interchange $\phi_+$ and $\phi_-$ in 
Eq.~(\ref{phipm}), we find the Dirac equation~(\ref{dirac}) does not change.
 Such solutions correspond to spin-down solutions.

The spin-down solution corresponding to the positive frequency mode
 $\Psi^+_{\uparrow}$ in Eq.~(\ref{up1}) is given by
\begin{eqnarray}
\Psi^{+}_{\downarrow}\left(z,\Omega\right)&=&
\left(
\begin{array}{l}
\varphi_{p}(z)\,\chi^{(+)}_{p\ell m}(\Omega)\,\\
\phi_{p}(z)\,\chi^{(+)}_{p\ell m}(\Omega)
\end{array}\right)\,.
\label{down1}
\end{eqnarray}
Note that the above solution is not only exchanging $\phi$ and $\varphi$ 
in Eq.~(\ref{up1}) but also $\chi_{p\ell m}^{(-)}$ changes to $\chi_{p\ell m}^{(+)}$.
Similary, the spin-down solution for the negative frequency mode 
$\Psi^-_{\uparrow}$ in Eq.~(\ref{up2}) is
\begin{eqnarray}
\Psi^-_{\downarrow}\left(z,\Omega\right)&=&\left(
\begin{array}{r}
-\varphi_{p}^*(z)\,\chi^{(-)}_{p\ell m}(\Omega)\,\\
\phi^*_{p}(z)\,\chi^{(-)}_{p\ell m}(\Omega)
\end{array}\right)\,.
\label{down2}
\end{eqnarray}

The general solution is expressed as a linear combination of the spin-up 
and down positive and negative frequency mode functions obtained in this section.

\section{Analytic Continuation}

In this section, we pick up the positive frequency mode functions 
corresponding to the Euclidean vacuum (the Bunch-Davies vacuum). 
To do this, we require that they are analytic when they are continued 
to the lower hemisphere of the Euclidean de Sitter space.

\subsection{The relation between the Lorentzian and the Euclidean coordinates}
The open chart is obtained by analytic continuation of the Euclidean sphere $S^4$.
The relation between the Lorentzian coordinate of three parts and the 
Euclidean coordinate is as follows.
\begin{eqnarray}
\left\{
\begin{array}{l}
t_R=i\left(\tau-\frac{\pi}{2}\right)\hspace{2.2cm}\,,t_R\geq 0\\
r_R=i\rho\hspace{3.5cm}\,,r_R\geq 0
\end{array}
\right.
\label{region:R}
\end{eqnarray}
\begin{eqnarray}
\left\{
\begin{array}{l}
t_C=\tau\hspace{2.5cm}\,,-\frac{\pi}{2}\leq t_C\leq \frac{\pi}{2}\\
r_C=i\left(\rho-\frac{\pi}{2}\right)\hspace{1.0cm}\,,0\leq r_C\leq \infty
\end{array}
\right.
\label{region:C}
\end{eqnarray}
\begin{eqnarray}
\left\{
\begin{array}{l}
t_L=i\left(-\tau-\frac{\pi}{2}\right)\hspace{2cm}\,,t_L\geq 0\\
r_L=i\rho\hspace{3.6cm}\,,r_L\geq 0
\end{array}
\right.
\end{eqnarray}

\subsection{The analytic continuation of the positive frequency mode function}
We first ignore the normalization and focus on the time dependent part 
of Eq.~(\ref{up1}) in region $R$ as
\begin{eqnarray}
\Psi^{+(R)}_\uparrow=\left(
\begin{array}{l}
\phi_{p}(z_R)\\
\varphi_{p}(z_R)
\end{array}\right)
\,,
\label{PsiPR1}
\end{eqnarray}
where the argument $z_R$ means that $\Psi^{+(R)}_\uparrow$ has support 
only in region $R$ and vanish in region $L$. In order to construct positive
 frequency mode functions corresponding to the Euclidean vacuum
 (the Bunch-Davies vacuum), we need to extend $\Psi^{+(R)}_\uparrow$ 
analytically from region $R$ to region $L$. The analytic continuation 
from the region $R$ to $L$ with the branch cut $[-1,1]$ goes through from 
Im$z<0$ to Im$z>0$.

To do it, we first extend $\Psi^{+(R)}_\uparrow$ analytically from region
 $R$ ($z_R\geq1$) to region $C$ ($-1\leq z_R\leq1$) by performing
 $z_R-1=e^{-\pi i}(1-z_R)$. Then we change the variable by $z_R=-z_L$ 
in the $C$ region. 
Since the hypergeometric function becomes singular at $z_L=1$,
 we use some formulae in Appendix~\ref{app:a}. Then we extend
 $\Psi^{+(R)}_\uparrow$ analytically from the region $C$ to 
region $L$ ($z_L>1$) by performing $1-z_L=e^{\pi i}(z_L-1)$. We then get
\begin{eqnarray}
\Psi^{+(R)}_\uparrow
&=&\left(
\begin{array}{r}
A\,\varphi_p(z_L)+B\,\phi_{p}^*(z_L)\,\\
-A\,\phi_p(z_L)+B\,\varphi_{p}^*(z_L)
\label{analytic1}
\end{array}\right)
\,,
\label{PsiPR2}
\end{eqnarray}
where we have defined
\begin{eqnarray}
A=\frac{H}{m}\,\frac{\Gamma(\frac{1}{2}-ip)\,
\Gamma(\frac{1}{2}+ip)}{\Gamma(-i\frac{m}{H})\,\Gamma(i\frac{m}{H})}
=\frac{\sinh\frac{m}{H}\pi}{\cosh\pi p}\,,\qquad
B=\frac{e^{-\pi p}\,
\Gamma(\frac{1}{2}-ip)^2}{\Gamma(\frac{1}{2}-ip-i\frac{m}{H})\,
\Gamma(\frac{1}{2}-ip+i\frac{m}{H})}\,.
\label{AB}
\end{eqnarray}
Note that Eq.~(\ref{PsiPR2}) has support only in region $L$ and 
vanishes in region $R$ due to the causally disconnected nature of
 region $R$ and $L$. Note also that for the case of a massless Dirac
 field $(m=0)$, $A=0$ and $B=e^{-\pi p}$. In the limit of 
$p=0$, $A=\sinh\frac{m}{H}\pi$ and $B=\cosh\frac{m}{H}\pi$.

From Eq.~(\ref{analytic1}), we find that the solution in the
 region $L$ can then be given by\footnote{One may think that the solution 
$\left(
\begin{array}{l}
\phi^*_{p}(z_L)\,\\
\varphi^*_{p}(z_L)
\end{array}\right)
$ could be another candidate solution in the region $L$. 
However, the solution does not satisfy the orthonormality relation.
 }
\begin{eqnarray}
\Psi^{+(L)}_\uparrow=\left(
\begin{array}{r}
\varphi_{p}(z_L)\,\\
-\phi_{p}(z_L)
\end{array}\right)
\,.
\label{up3}
\end{eqnarray}
where the argument $z_L$ means $\Psi^{+(L)}_\uparrow$ has support only
 in region $L$ and vanishes in region $R$.
Thus if we extend $\Psi^{+(L)}_\uparrow$ analytically from region 
$L$ to region $R$, the above solution becomes
\begin{eqnarray}
\Psi^{+(L)}_\uparrow=
&=&\left(
\begin{array}{r}
-A\,\phi_p(z_R)+B\,\varphi_{p}^*(z_R)\,\\
-A\,\varphi_p(z_R)-B\,\phi_{p}^*(z_R)
\end{array}\right)
\,.
\label{PsiPL2}
\end{eqnarray}

By using the solutions given in Eqs.~(\ref{PsiPR1}), (\ref{PsiPR2}), (\ref{up3})
 and (\ref{PsiPL2}),  we find that the above solution satisfies the 
orthonormality relation
\begin{eqnarray}
\left(\Psi_{p}^{+(R)},\Psi_{p'}^{+(L)}\right)=-A+A^*=0\,,
\end{eqnarray}
in the far past $z\rightarrow 1$.

\subsection{The analytic continuation of the negative frequency mode function}
We write the time dependent part of Eq.~(\ref{up2}) as
\begin{eqnarray}
\Psi^{-(R)}_\uparrow=\left(
\begin{array}{r}
\phi^*_{p}(z_R)\\
-\varphi^*_{p}(z_R)
\end{array}\right)\,.
\label{PsiMR}
\end{eqnarray}
We want to extend $\Psi^{-(R)}_\uparrow$ analytically from region $R$ 
to region $L$. The analytic continuation from the region $R$ to $L$ 
with the branch cut $[-1,1]$ goes through from Im$z>0$ to Im$z<0$. 
The procedure is the same as what we did for the positive frequency mode. 
The solution extended analytically from the region $R$ to $L$ is then give by
\begin{eqnarray}
\Psi^{-(R)}_\uparrow
&=&\left(
\begin{array}{r}
A^*\varphi_p^*(z_L)+B^*\phi_{p}(z_L)\,\\
A^*\phi_p^*(z_L)-B^*\varphi_{p}(z_L)
\label{analytic2}
\end{array}\right)\,,
\end{eqnarray}
where $A$ and $B$ are given in Eq.~(\ref{AB}).

The solution in the region $L$ then can become
\begin{eqnarray}
\Psi^{-(L)}_\uparrow=\left(
\begin{array}{c}
\varphi^*_{p}(z_L)\,\\
\phi^*_{p}(z_L)
\end{array}\right)
\,.
\label{up4}
\end{eqnarray}
If we extend $\Psi^{-(L)}_\uparrow$ from region $L$ to $R$ analytically, 
the above solution becomes
\begin{eqnarray}
\Psi^{-(L)}_\uparrow=
&=&\left(
\begin{array}{r}
-A^*\phi^*_p(z_R)+B^*\varphi_{p}(z_R)\,\\
A^*\varphi^*_p(z_R)+B^*\phi_{p}(z_R)
\end{array}\right)
\,.
\end{eqnarray}
The above solution satisfies the orthonormality relation
\begin{eqnarray}
\left(\Psi_{p}^{+(R)},\Psi_{p'}^{+(L)}\right)=-A+A^*=0\,,
\end{eqnarray}
in the far past $z\rightarrow 1$.

\section{The Bunch-Davies vacuum solutions}

In the previous section, we found that mode functions obtained in one region 
are analytically continued to the other region by
 $\phi\rightarrow A\varphi+B\phi^*$ and $\varphi\rightarrow -A\phi+B\varphi^*$.
 In this section, we present the Bunch-Davies vacuum solutions of positive 
and negative frequency mode functions.

\subsection{Positive frequency mode}
The positive frequency mode functions that have support on both region $R$ 
and $L$ are found to be
\begin{eqnarray}
\Psi^{+(R)}_{\uparrow}(z,\Omega)=\left\{
\begin{array}{l}
\frac{1}{N_b}\left(
\begin{array}{r}
\phi_{p}(z)\,\chi^{(-)}_{p\ell m}(\Omega)\\
\varphi_{p}(z)\,\chi^{(-)}_{p\ell m}(\Omega)
\end{array}\right)\hspace{3.9cm} {\rm for}\quad z=z_R\,,\\\\
\frac{1}{N_b}\left(
\begin{array}{r}
\left(A\,\varphi_p(z)+B\,\phi_{p}^*(z)\,\right)\chi^{(-)}_{p\ell m}(\Omega)\\
\left(-A\,\phi_p(z)+B\,\varphi_{p}^*(z)\,\right)\chi^{(-)}_{p\ell m}(\Omega)
\end{array}\right)\qquad {\rm for}\quad z=z_L\,,\\
\end{array}\right.\\\nonumber\\\nonumber\\
\Psi^{+(R)}_{\downarrow}(z,\Omega)=\left\{
\begin{array}{l}
\frac{1}{N_b}\left(
\begin{array}{l}
\varphi_{p}(z)\,\chi^{(+)}_{p\ell m}(\Omega)\\
\phi_{p}(z)\,\chi^{(+)}_{p\ell m}(\Omega)
\end{array}\right)\hspace{3.9cm} {\rm for}\quad z=z_R\,,\\\\
\frac{1}{N_b}\left(
\begin{array}{r}
\left(-A\,\phi_p(z)+B\,\varphi_{p}^*(z)\,\right)\chi^{(+)}_{p\ell m}(\Omega)\\
\left(A\,\varphi_p(z)+B\,\phi_{p}^*(z)\,\right)\chi^{(+)}_{p\ell m}(\Omega)
\end{array}\right)\qquad {\rm for}\quad z=z_L\,,\\
\end{array}\right.
\end{eqnarray}
\begin{eqnarray}
\Psi^{+(L)}_{\uparrow}(z,\Omega)=\left\{
\begin{array}{l}
\frac{1}{N_b}\left(
\begin{array}{l}
\left(-A\,\phi_p(z)+B\,\varphi_{p}^*(z)\,\right)\chi^{(-)}_{p\ell m}(\Omega)\\
-\left(A\,\varphi_p(z)+B\,\phi_{p}^*(z)\,\right)\chi^{(-)}_{p\ell m}(\Omega)
\end{array}\right)\hspace{0.8cm} {\rm for}\quad z=z_R\,,\\\\
\frac{1}{N_b}\left(
\begin{array}{r}
\varphi_{p}(z)\,\chi^{(-)}_{p\ell m}(\Omega)\,\\
-\phi_{p}(z)\,\chi^{(-)}_{p\ell m}(\Omega)
\end{array}\right)\hspace{3.6cm} {\rm for}\quad z=z_L\,.
\end{array}\right.\\\nonumber\\\nonumber\\
\Psi^{+(L)}_{\downarrow}(z,\Omega)=\left\{
\begin{array}{l}
\frac{1}{N_b}\left(
\begin{array}{l}
-\left(A\,\varphi_p(z)+B\,\phi_{p}^*(z)\,\right)\chi^{(+)}_{p\ell m}(\Omega)\\
\left(-A\,\phi_p(z)+B\,\varphi_{p}^*(z)\,\right)\chi^{(+)}_{p\ell m}(\Omega)
\end{array}\right)\hspace{0.8cm} {\rm for}\quad z=z_R\,,\\\\
\frac{1}{N_b}\left(
\begin{array}{r}
-\phi_{p}(z)\,\chi^{(+)}_{p\ell m}(\Omega)\,\\
\varphi_{p}(z)\,\chi^{(+)}_{p\ell m}(\Omega)
\end{array}\right)
\label{down3}
\hspace{3.6cm} {\rm for}\quad z=z_L\,.
\end{array}\right.
\end{eqnarray}
The normalization factor $N_b$ is computed by 
$\sqrt{-g}\,\Psi_{p}^{+(R)\dag}\Psi_{p}^{+(R)}=1$ at $z\rightarrow 1$, 
which is expressed as
\begin{eqnarray}
N_b^2=1+\frac{\sinh^2\pi\frac{m}{H}}{\cosh^2\pi p} 
+\frac{\pi^2\,e^{-2\pi p}}{\cosh^2\pi p}
\frac{1}{|\Gamma(\frac{1}{2}-ip-i\frac{m}{H})\,
\Gamma(\frac{1}{2}-ip+i\frac{m}{H})|^2}\,.
\label{nb}
\end{eqnarray}
Note that for a massless Dirac field, $N_b=1+e^{-2\pi p}$.

\subsection{Negative frequency mode}
The negative frequency mode functions that have support on both
 region $R$ and $L$ are found to be
\begin{eqnarray}
\Psi^{-(R)}_{\uparrow}(z,\Omega)=\left\{
\begin{array}{l}
\frac{1}{N_b}\left(
\begin{array}{r}
\phi^*_{p}(z)\,\chi^{(+)}_{p\ell m}(\Omega)\\
-\varphi^*_{p}(z)\,\chi^{(+)}_{p\ell m}(\Omega)
\end{array}\right)\hspace{3.8cm} {\rm for}\quad z=z_R\,,\\\\
\frac{1}{N_b}\left(
\begin{array}{c}
\left(A^*\varphi^*_p(z)+B^*\phi_{p}(z)\,\right)\chi^{(+)}_{p\ell m}(\Omega)\\
\left(A^*\phi^*_p(z)-B^*\varphi_{p}(z)\,\right)\chi^{(+)}_{p\ell m}(\Omega)
\end{array}\right)\hspace{1.2cm} {\rm for}\quad z=z_L\,,\\
\end{array}\right.\\\nonumber\\\nonumber\\
\Psi^{-(R)}_{\downarrow}(z,\Omega)=\left\{
\begin{array}{l}
\frac{1}{N_b}\left(
\begin{array}{r}
-\varphi^*_{p}(z)\,\chi^{(-)}_{p\ell m}(\Omega)\\
\phi^*_{p}(z)\,\chi^{(-)}_{p\ell m}(\Omega)
\end{array}\right)\hspace{3.8cm} {\rm for}\quad z=z_R\,,\\\\
\frac{1}{N_b}\left(
\begin{array}{c}
\left(A^*\phi^*_p(z)-B^*\varphi_{p}(z)\,\right)\chi^{(-)}_{p\ell m}(\Omega)\\
\left(A^*\varphi^*_p(z)+B^*\phi_{p}(z)\,\right)\chi^{(-)}_{p\ell m}(\Omega)
\end{array}\right)\hspace{1.2cm} {\rm for}\quad z=z_L\,,\\
\end{array}\right.
\end{eqnarray}
\begin{eqnarray}
\Psi^{-(L)}_{\uparrow}(z,\Omega)=\left\{
\begin{array}{l}
\frac{1}{N_b}\left(
\begin{array}{r}
\left(-A^*\phi^*_p(z)+B^*\varphi_{p}(z)\,\right)\chi^{(+)}_{p\ell m}(\Omega)\\
\left(A^*\varphi^*_p(z)+B^*\phi_{p}(z)\,\right)\chi^{(+)}_{p\ell m}(\Omega)
\end{array}\right)\hspace{1.0cm} {\rm for}\quad z=z_R\,,\\\\
\frac{1}{N_b}\left(
\begin{array}{c}
\varphi^*_{p}(z)\,\chi^{(+)}_{p\ell m}(\Omega)\,\\
\phi^*_{p}(z)\,\chi^{(+)}_{p\ell m}(\Omega)
\end{array}\right)\hspace{4.2cm} {\rm for}\quad z=z_L\,.
\end{array}\right.\\\nonumber\\\nonumber\\
\Psi^{-(L)}_{\downarrow}(z,\Omega)=\left\{
\begin{array}{l}
\frac{1}{N_b}\left(
\begin{array}{r}
\left(A^*\varphi^*_p(z)+B^*\phi_{p}(z)\,\right)\chi^{(-)}_{p\ell m}(\Omega)\\
\left(-A^*\phi^*_p(z)+B^*\varphi_{p}(z)\,\right)\chi^{(-)}_{p\ell m}(\Omega)
\end{array}\right)\hspace{1.0cm} {\rm for}\quad z=z_R\,,\\\\
\frac{1}{N_b}\left(
\begin{array}{c}
\phi^*_{p}(z)\,\chi^{(-)}_{p\ell m}(\Omega)\,\\
\varphi^*_{p}(z)\,\chi^{(-)}_{p\ell m}(\Omega)
\end{array}\right)
\label{down4}
\hspace{4.2cm} {\rm for}\quad z=z_L\,.
\end{array}\right.
\end{eqnarray}
The normalization factor $N_b$ is given by Eq.~(\ref{nb}).

\section{Supercurvature modes}

In this section, we examine if there is a supercurvature mode for
 the Dirac field. The supercurvature modes exist in the open chart for
 a massive scalar field~\cite{Sasaki:1994yt}. They are known to be able
 to carry information about the pre-tunneling vacuum state and to be
 expected that they may explain the dipolar statistical 
anisotropy~\cite{Liddle:2013czu, Kanno:2013ohv, Kobayashi:2015qma}.
 For a massive vector field, the paper~\cite{Yamauchi:2014saa} concluded 
that there is no supercurvature mode in the open chart.
 
We examine if the mode functions (\ref{phi}) and (\ref{varphi}) for 
$p=i\tilde{p}$ is normalizable in $C$ region (see Figure~\ref{fig1}).
The metric in the $C$ region is
\begin{eqnarray}
ds^2=H^{-2}\left[\,dt_C^2
+\cos^2t_C\left(-dr^2_C+\cosh^2r_C\,d\Omega^2\right)\,\right]\,.
\end{eqnarray}
Note that the volume element is given by $\sqrt{h}=H^{-3}\cos^2t_C\cosh^2r_C$.
By using Eqs.~(\ref{region:R}) and (\ref{region:C}), we find the relation 
$z_R=\cosh t_R=\sin t_C$, and $z_R=\pm 1$ correspond to $t_C=\pm\pi/2$. 
The mode function $\Psi^{\dag (R)}_\uparrow$ becomes real in the $C$ region 
except for the irrelevant overall phase when $p$ is pure imaginary.
Since the hypergeometric functions in Eqs.~(\ref{phi}) and (\ref{varphi}) 
converge to unity at $z=1$, we can easily read the asymptotic behavior of 
the mode function 
$\Psi^{\dag(R)}_\uparrow$ at $t_C=\pm\pi/2$ from 
Eqs.~\eqref{PsiPR1} and \eqref{PsiPR2}. 
If we focus on the behavior around $t_C=\pi/2$, {\it i.e.} $z_R=1$, 
the more  
singular is the upper component proportional to 
$\phi_p(\sin t_C)\sim
\left(z_R-1\right)^{-\frac{3}{4}-\frac{ip}{2}}
=\left(\sin t_C-1\right)^{-\frac{3}{4}-\frac{ip}{2}}
$.  
Since the normalization is given by 
\begin{eqnarray}
\int_{-\frac{\pi}{2}}^{\frac{\pi}{2}} dt_C\sqrt{h}\,
\Psi_p^{+(R)\dag}\Psi_p^{+(R)}\sim
\int_{-\frac{\pi}{2}}^{\frac{\pi}{2}} dt_C\sqrt{h}\,
\left[\,\phi_p^2(\sin t_C)+\varphi_p^2(\sin t_C)\,\right]
\,, 
\end{eqnarray}
the contribution from the vicinity of $t_C=\pi/2$ converges if $ip$ is negative. 
The behavior on the other boundary of the region $C$ can be read from \eqref{PsiPR2}. 
For $ip<0$, the most singular piece in 
\eqref{PsiPR2} at $t_C=-\pi/2$ is proportional to $B\phi^*_p(-\sin t_C)$. 
Thus as long as $B$ does not vanish, the above integral 
that determines the normalization diverges. Since $B$ never vanishes
as easily seen from Eq.~\eqref{AB},
 we conclude that there is no normalizable mode for $p^2<0$. 
Thus there exists no supercurvature mode.

\section{Entanglement entropy}

In this section, we quantize the Dirac field in the open chart in order to 
discuss quantum entanglement. In order to discuss quantum entanglement from
 the point of view of, say $R$ region, we need to trace out the degree of 
freedom of inaccessible $L$ region. Thus we need to change basis to mode 
functions that have support on either $R$ or $L$ regions. Thus, we consider
 the Bogoliubov transformation between the Bunch-Davies vacuum and $R$, $L$ vacua.
 We then derive the reduced density matrix in the region $R$ by tracing
 over the region $L$ and compute the entanglement entropy with the reduced density matrix.

\subsection{Canonical quantization}

If we take into account the two sets of modes in each $R$ and $L$ region, 
the Dirac field can be expanded in terms of the spin-up and down positive 
and negative frequency mode $\Psi^{+(R)/(L)}_s$ and $\Psi^{-(R)/(L)}_s$ 
respectively such as
\begin{eqnarray}
\Psi=\int dp\sum_{\ell m}\sum_s
\left(\,a_{s}^R\,\Psi^{+(R)}_s+b_{s}^{R\dag}\,\Psi^{-(R)}_s
+a_{s}^L\,\Psi^{+(L)}_s+b_{s}^{L\dag}\,\Psi^{-(L)}_s\,\right)\,,
\label{expansion1}
\end{eqnarray}
where $s=(\uparrow,\downarrow)$. The operators $a_{s}^{R\dag}$, 
$b_{s}^{R\dag}$ and $a_{s}^R$, $b_{s}^{R}$ are the creation and annihilation 
operators for the positive and negative frequency modes that satisfy 
the anticommutation relations
\begin{eqnarray}
\left\{a_{sp\ell m}^\sigma,~a_{s'p'\ell' m'}^{\sigma'\dag}\right\}
=\left\{b_{sp\ell m}^\sigma,~b_{s'p'\ell' m'}^{\sigma'\dag}\right\}
=\delta\left(p-p'\right)\delta_{ss'}\delta_{\ell\ell'}\delta_{mm'}
\delta_{\sigma\sigma'}\,,
\label{anticommutation1}
\end{eqnarray}
where $\sigma=R$ or $L$ and we recovered abbreviated subscripts such as $p,\ell, m$ temporally but they are omitted below for simplicity unless there may be any confusion. All other anticommutators vanish. If we focus 
on a mode $p$, the Bunch-Davies vacuum is defined by
\begin{eqnarray}
|0\rangle_{\rm BD}=|0\rangle^+_{\rm BD}|0\rangle^-_{\rm BD}\,,
\end{eqnarray}
where the $+$ superscript on the ket indicates the spin-down particle 
and spin-up antiparticle vacua and the $-$ superscript for the spin-up 
particle and spin-down antiparticle vacua respectively, so that
\begin{eqnarray}
&&a^\sigma_\downarrow|0\rangle^+_{\rm BD}=b^\sigma_\uparrow|0\rangle^+_{\rm BD}=0\,,
\end{eqnarray}
where $\sigma=R$ and $L$. Since $(a_s^\dag)^2=(b_s^\dag)^2=0$, 
only two states are allowed for particles and antiparticles.

\subsection{The Bogoliubov transformation}

We denote the spin-up (\ref{up1}) and down (\ref{down1}) positive 
frequency mode functions in region $R$  by $\psi^{+(R)}_\uparrow$ 
and $\psi^{+(R)}_\downarrow$ such as
\begin{eqnarray}
\psi^{+(R)}_\uparrow=
\left(
\begin{array}{l}
\phi_{p}(z_R)\\
\varphi_{p}(z_R)
\end{array}\right)
\chi^{(-)}_{p\ell m}(\Omega)\,,\qquad
\psi^{+(R)}_\downarrow=
\left(
\begin{array}{l}
\varphi_{p}(z_R)\\
\phi_{p}(z_R)
\end{array}\right)
\chi^{(+)}_{p\ell m}(\Omega)\,.
\end{eqnarray}
Similarly for the spin-up (\ref{up3}) and down (\ref{down3}) mode
 functions in region $L$,
\begin{eqnarray}
\psi^{+(L)}_\uparrow=
\left(
\begin{array}{r}
\varphi_{p}(z_L)\\
-\phi_{p}(z_L)
\end{array}\right)
\chi^{(-)}_{p\ell m}(\Omega)\,,\qquad
\psi^{+(L)}_\downarrow=
\left(
\begin{array}{r}
-\phi_{p}(z_L)\\
\varphi_{p}(z_L)
\end{array}\right)
\chi^{(+)}_{p\ell m}(\Omega)\,.
\end{eqnarray}
For the negative frequency mode functions, we express the 
spin-up (\ref{up2}) and down (\ref{down2}) solutions by
\begin{eqnarray}
\psi^{-(R)}_\uparrow=
\left(
\begin{array}{r}
\phi^*_{p}(z_R)\\
-\varphi^*_{p}(z_R)
\end{array}\right)
\chi^{(+)}_{p\ell m}(\Omega)\,,\qquad
\psi^{-(R)}_\downarrow=
\left(
\begin{array}{r}
-\varphi^*_{p}(z_R)\\
\phi^*_{p}(z_R)
\end{array}\right)
\chi^{(-)}_{p\ell m}(\Omega)\,,
\end{eqnarray}
and for the spin-up (\ref{up4}) and down (\ref{down4}) negative 
frequency mode functions in region $L$ we write
\begin{eqnarray}
\psi^{-(L)}_\uparrow=
\left(
\begin{array}{r}
\varphi^*_{p}(z_L)\\
\phi^*_{p}(z_L)
\end{array}\right)
\chi^{(-)}_{p\ell m}(\Omega)\,,\qquad
\psi^{-(L)}_\downarrow=
\left(
\begin{array}{r}
\phi^*_{p}(z_L)\\
\varphi^*_{p}(z_L)
\end{array}\right)
\chi^{(+)}_{p\ell m}(\Omega)\,.
\end{eqnarray}

In region $R$, let us denote $(c_s^R\,,c_s^{R\dag})$ as the annihilation 
and creation operators for fermions and $(d_s^R\,,d_s^{R\dag})$ as the 
annihilation and creation operators for antifermions. The corresponding 
fermion and antifermion operators in region $L$ are denoted as 
$(c_s^L\,,c_s^{L\dag})$ and $(d_s^L\,,d_s^{L\dag})$. 
Then the Dirac field can be expanded as
\begin{eqnarray}
\Psi=\int dp\sum_{\ell m}
\sum_s\left(\,c_{s}^R\,\psi^{+(R)}_s+d_{s}^{R\dag}\,\psi^{-(R)}_s
+c_{s}^L\,\psi^{+(L)}_s+d_{s}^{L\dag}\,\psi^{-(L)}_s\,\right)\,.
\label{expansion2}
\end{eqnarray}
These obey the anticommutation relations
\begin{eqnarray}
\left\{c_{sp\ell m}^\sigma,~c_{s'p'\ell' m'}^{\sigma'\dag}\right\}
=\left\{d_{sp\ell m}^\sigma,~d_{s'p'\ell' m'}^{\sigma'\dag}\right\}
=\delta\left(p-p'\right)\delta_{ss'}\delta_{\ell\ell'}\delta_{mm'}
\delta_{\sigma\sigma'}\,,
\end{eqnarray}
where $\sigma=R$ or $L$. All other anticommutators 
vanish~\footnote{The indices $p,\ell,m$ are omitted below for simplicity
 unless there may be any confusion.}. 

As the Dirac field operator should be the same under this change of basis,
  we can relate the creation and annihilation operators in the Bunch-Davies
 vacuum to those in $R, L$ vacua by comparing Eq.~(\ref{expansion1}) 
with (\ref{expansion2}) It follows that
\begin{eqnarray}
\left(
c^R_\downarrow\,,\,
c^L_\downarrow\,,\,
d^{R\dag}_\uparrow\,,\,
d^{L\dag}_\uparrow
\right)=
\left(
a^R_\downarrow\,,\,
a^L_\downarrow\,,\,
b^{R\dag}_\uparrow\,,\,
b^{L\dag}_\uparrow
\right){\bf M}\,,
\label{cdab}
\end{eqnarray}
We have another set of four relations with their spin-up and
 down exchanged but it is totally equivalent.
Thus, we focus on the relation (\ref{cdab}) below.
 Here  ${\bf M}$ is a $4\times 4$ matrix
\begin{eqnarray}
{\bf M}=\left(
\begin{array}{cc}
\alpha & \beta\\
-\beta^{*} & \alpha^{*} \\
\end{array}\right)\,,
\end{eqnarray}
where $\alpha$ and $\beta$ are $2\times 2$ matrices respectively defined as
\begin{eqnarray}
\alpha=\frac{1}{N_b}\left(
\begin{array}{cc}
1 & -A \\
A & 1 \\
\end{array}\right)\,,\qquad
\beta=\frac{1}{N_b}\left(
\begin{array}{cc}
0 & -B \\
B & 0 \\
\end{array}\right)\,.
\label{alphabeta}
\end{eqnarray}
Note that $A, B$ and $N_b$ are given in Eqs.~(\ref{AB}) and (\ref{nb}).  
Then, we find
\begin{eqnarray}
\left(
a^{R}_\downarrow\,,\,
a^{L}_\downarrow\,,\,
b^{R\dag}_{\uparrow}\,,\,
b^{L\dag}_{\uparrow}
\right)=
\left(
c^R_\downarrow\,,\,
c^L_\downarrow\,,\,
d^{R\dag}_{\uparrow}\,,\,
d^{L\dag}_{\uparrow}
\right){\bf M}^{-1}\,,
\label{abcd}
\end{eqnarray}
where
\begin{eqnarray}
{\bf M}^{-1}=
\left(
\begin{array}{cc}
\xi & \delta \\
-\delta^* & \xi^* \\
\end{array}\right)\,,
\end{eqnarray}
and the components of the matrix ${\bf M}^{-1}$ are computed as
\begin{eqnarray}
\xi=\left(\alpha+\beta\alpha^{*-1}\beta^*\right)^{-1}\,,\qquad
\delta=-\alpha^{-1}\beta\xi^*\,.
\label{gammadelta}
\end{eqnarray}

Since $a_\downarrow^\sigma$ mixes the spin-down particles and the spin-up
 antiparticles in region $R$ and $L$, the Bunch-Davies vacuum for mode $p$ 
can be regarded as the Bogoliubov transformation of $R$, $L$ vacua of the form
\begin{eqnarray}
|0\rangle^+_{\rm BD}\propto
\exp\left(\,\sum_{i,j=R,L}m_{ij}\,c^{i\dagger} d^{j\dagger}\right)
 |0\rangle_R^+|0\rangle_L^-\,,
\label{bogoliubov1}
\end{eqnarray}
where $m_{ij}$ is an antisymmetric matrix and operators $c^i$ and $d^j$ 
satisfy the anticommutation relation $\{c^i,d^j\}=0$. 
Here the normalization of the Bogoliubov transformation is omitted because
 we will need another Bogoliubov transformation in Eq.~(\ref{bogoliubov2}). 
Note that we drop spin labels on operators for simplicity because we focus
 on operators $c_\downarrow$ and $d_\uparrow$ as defined by
\begin{eqnarray}
c_\downarrow^R\,|0\rangle^+_{R}=d_\uparrow^R\,|0\rangle^+_{R}=0\,,\qquad
c_\downarrow^L|0\rangle^-_{L}=d_\uparrow^L|0\rangle^-_{L}=0\,.
\end{eqnarray}

If we apply $a_\downarrow^{\sigma}$ in Eq.~(\ref{abcd}) to
 Eq.~(\ref{bogoliubov1}), we have
\begin{eqnarray}
0=a_\downarrow^\sigma\,|0\rangle^+_{\rm BD}\Longrightarrow
m_{ij}=\left(\delta^*\xi^{-1}\right)_{ij}\,,
\end{eqnarray}
where $\sigma=R$ and $L$, and we used the anticommutation 
relation Eq.~(\ref{anticommutation1}).
Using the expressions in Eqs.~(\ref{alphabeta}) and (\ref{gammadelta}), 
we find
\begin{eqnarray}
m_{ij}=-\frac{B^*}{1+A^2}
\left(
\begin{array}{cc}
A & -1 \vspace{3mm}\\
1 & ~A \\
\end{array}\right)
\,,
\label{mij}
\end{eqnarray}
where $A$ and $B$ are given in Eq.~(\ref{AB}). Since $A=0$ in the case 
of masslessness, the density matrix 
$\rho=|0\rangle^+_{\rm BD}{}^{~+}_{\rm BD}\langle 0|$ in terms of
 Eq.~(\ref{bogoliubov1}) becomes diagonal in the $|0\rangle_R^+|0\rangle_L^-$
 basis. In other cases, however, it is not diagonal and then
 it is difficult to trace out the $L$ degrees of freedom. 
Thus, we introduce new operators $\tilde{c}_R$ and $\tilde{d}_L$ 
and perform a further Bogoliubov transformation
\begin{eqnarray}
\tilde{c}_R&=&u\,c_R+v\,d_R^\dag\,,\qquad
\tilde{d}_L=u^*d_L+v^*c_L^\dag\,,\nonumber\\
\tilde{d}_R&=&u\,d_R-v\,c_R^\dag\,,\qquad
\tilde{c}_L=u^*c_L-v^*d_L^\dag\,,
\label{tilde}
\end{eqnarray}
with $|u|^2+|v|^2=1$, so that we obtain the form
\begin{eqnarray}
|0\rangle^+_{\rm BD}=N_{\gamma_p}^{-1}\exp\left(\gamma_p\,\tilde{c}_R^\dagger\, \tilde{d}_L^\dagger+\gamma_p\,\tilde{d}_R^\dagger\, \tilde{c}_L^\dagger\right) |0\rangle_{R'}^+|0\rangle_{L'}^-\,,
\label{bogoliubov2}
\end{eqnarray}
where $|0\rangle_{R'}^+=|0_c\rangle_{R'}^+|0_d\rangle_{R'}^+$ and $|0\rangle_{L'}^-=|0_c\rangle_{L'}^-|0_d\rangle_{L'}^-$. The subscripts $c$ and $d$ in the kets are used to indicate the vacua annihilated by the operators $\tilde{c}_\sigma$ and $\tilde{d}_\sigma$, respectively. Note that the Bogoliubov transformation in Eq.~(\ref{tilde}) does not mix $R$ and $L$ Hilbert spaces because Eq.~(\ref{tilde}) is a linear transformation between $\tilde{c}_\sigma, \tilde{d}_\sigma$ and $c_\sigma, d_\sigma$.
The new operators satisfy the anticommutation relation $\{\tilde{c}_i,\tilde{d}_j^\dag\}=\delta_{ij}$. Normalization determines the $N^2_{\gamma_p}$
\begin{eqnarray}
{}_{\rm BD}^{~\,+}\langle 0|0\rangle^+_{\rm BD}=N_{\gamma_p}^2\left(1+|\gamma_p|^2\right)=1\,,
\end{eqnarray}
where $|\gamma_p|<1$ should be satisfied. If we use the anticommutation relations $\{\tilde{c}_i,\tilde{c}_j^\dag\}=\delta_{ij}$ and $\{\tilde{d}_i,\tilde{d}_j^\dag\}=\delta_{ij}$, Eq.~(\ref{bogoliubov2}) gives the consistency conditions
\begin{eqnarray}
&&\tilde{c}_R|0\rangle^+_{\rm BD}=\gamma_p\,\tilde{d}_L^\dag|0\rangle^+_{\rm BD}\,,\qquad
\tilde{d}_L|0\rangle^+_{\rm BD}=-\gamma_p\,\tilde{c}_R^\dag|0\rangle^+_{\rm BD}\,,
\label{consistency1}\\
&&\tilde{d}_R|0\rangle^+_{\rm BD}=\gamma_p\,\tilde{c}_L^\dag|0\rangle^+_{\rm BD}\,,\qquad
\tilde{c}_L|0\rangle^+_{\rm BD}=-\gamma_p\,\tilde{d}_R^\dag|0\rangle^+_{\rm BD}\,.
\label{consistency2}
\end{eqnarray}
Let us write $m_{RR}=m_{LL}\equiv\omega$ and $m_{RL}=-m_{LR}=\zeta$ in Eq.~(\ref{mij}). Then, the first condition of Eq.~(\ref{consistency1}) imposes constraints on $u$ and $v$
\begin{eqnarray}
u\omega+v+\gamma_p v\zeta=0\,,\qquad
u\zeta-\gamma_p u-\gamma_p v\omega=0\,,\qquad
\label{uv}
\end{eqnarray}
The first relation in Eq.~(\ref{uv}) gives
$u/v=-\left(\gamma_p\zeta+1\right)/\omega$. Plugging this into the second relation in Eq.~(\ref{uv}), we obtain
\begin{eqnarray}
\gamma_p=\frac{1}{2\zeta}\left[\,
\omega^2+\zeta^2-1+\sqrt{\left(\omega^2+\zeta^2-1\right)^2+4\zeta^2}\,\right]\,,
\label{gammap}
\end{eqnarray}
where a plus sign in front of the square root term is taken to satisfy $|\gamma_p|<1$.
Note that the second condition of Eq.~(\ref{consistency1}) leads to the relations for $u^*/v^*$. We find $u^*/v^*=-u/v$ and we obtain the same $\gamma_p$ in Eq.~(\ref{gammap}). The condition Eq.~(\ref{consistency2}) gives the same $\gamma_p$ in Eq.~(\ref{gammap}) and the relation $u^*/v^*=-u/v$ as well.
Note also that in the limit of $p=0$, $\omega=-A/B$, $\zeta=1/B$ and $\omega^2+\zeta^2=1$ holds, thus we find $\gamma_p=1$ independently of $m$.

\subsection{The density matrix}

\begin{figure}[t]
\vspace{-2cm}
\includegraphics[height=10cm]{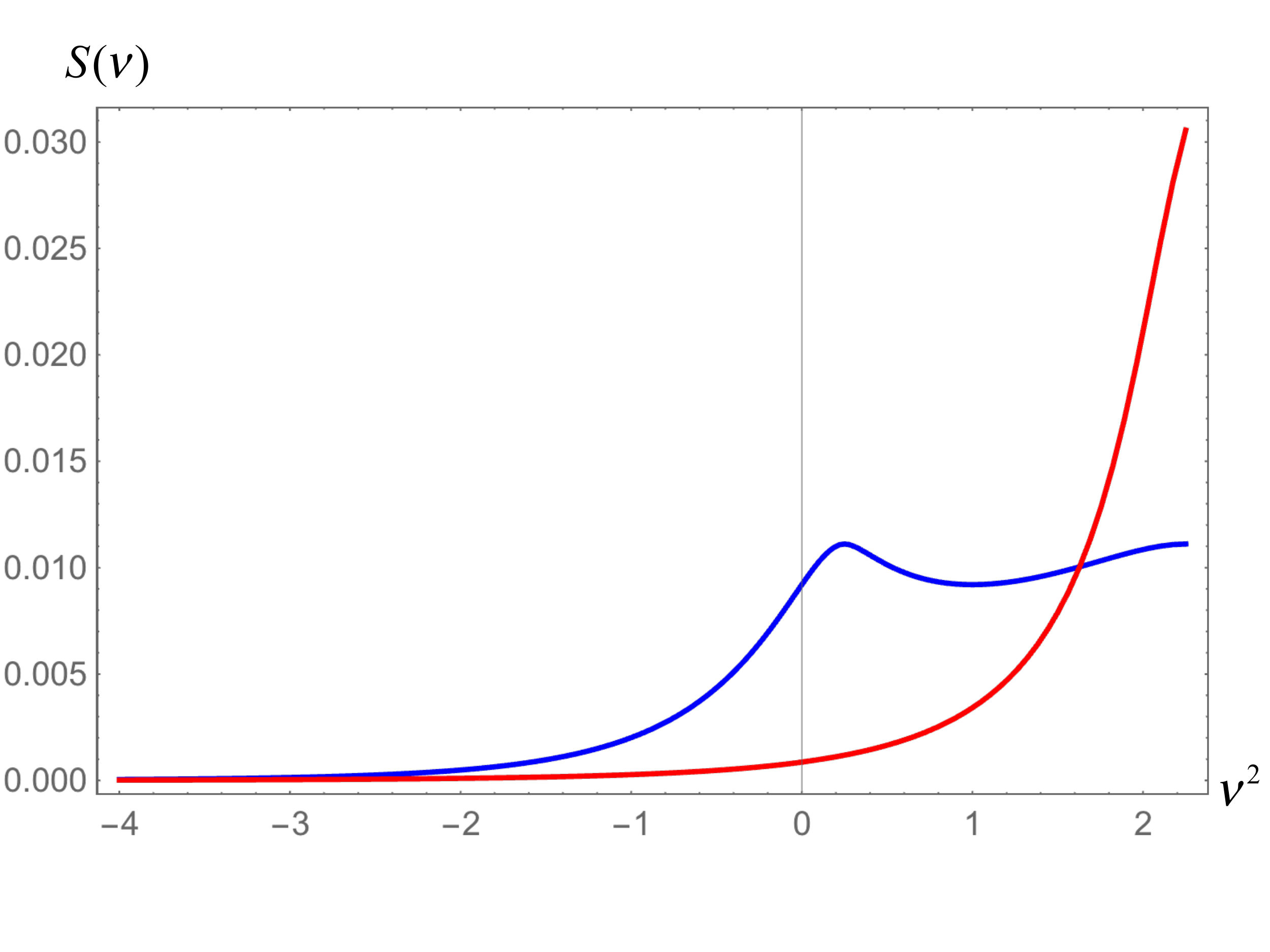}\centering
\vspace{-1.5cm}
\caption{Entanglement entropy between two causally disconnected regions for one degree of freedom of a fermion (red) and a scalar field (blue).}
\label{fig2}
\end{figure}

By using Eqs.~(\ref{bogoliubov2}) and (\ref{gammap}), the reduced density matrix in region $R$ is then found to be
\begin{eqnarray}
\rho_R&=&{\rm Tr}_L\,|0\rangle^+_{\rm BD}\,{}_{\rm BD}^{~\,+}\langle 0|\nonumber\\
&=&\frac{1}{\left(1+|\gamma_p|^2\right)^2}\Bigl(
|0_c\rangle^+_{R'}\,{}_{R'}^{+}\langle 0_c|
+|\gamma_p|^2|1_c\rangle^+_{R'}\,{}_{R'}^{+}\langle 1_c|
\Bigr)\nonumber\\
&&\hspace{2.5cm}\otimes
\Bigl(
|0_d\rangle^+_{R'}\,{}_{R'}^{+}\langle 0_d|
+|\gamma_p|^2|1_d\rangle^+_{R'}\,{}_{R'}^{+}\langle 1_d|
\Bigr)
\,,
\end{eqnarray}
where the conservation of probability holds, ${\rm Tr}\rho_R=1$. 

As an observer in the region $R$ will observe particles defined by the operators $\tilde{c}_R$, the expected number of such particles will be given by
\begin{eqnarray}
{}_{\rm BD}^{~\,+}\langle 0|\tilde{c}_R^\dag\tilde{c}_R|0\rangle^+_{\rm BD}
&=&{\rm Tr}_R\,\tilde{c}_R^\dag\tilde{c}_R\,\rho_R\nonumber\\
&=&\frac{|\gamma_p|^2}{\left(1+|\gamma_p|^2\right)^2}\,
{}_{R'}^{+}\langle 1_c|\tilde{c}_R^\dag\tilde{c}_R|1_c\rangle^+_{R'}
+\frac{|\gamma_p|^4}{\left(1+|\gamma_p|^2\right)^2}\,
{}_{R'}^{+}\langle 1_d\,1_c|\tilde{c}_R^\dag\tilde{c}_R|1_c\,1_d\rangle^+_{R'}
\nonumber\\
&=&\frac{1}{|\gamma_p|^{-2}+1}
\,,
\end{eqnarray}
where for notational convenience we have defined $|1_c\,1_d\rangle_{R'}^+=|1_c\rangle_{R'}^+|1_d\rangle_{R'}^+$. In the case of masslessness, this is expressed by $\left(e^{2\pi p}+1\right)^{-1}$, which is a thermal state with temperature
\begin{eqnarray}
T=\frac{H}{2\pi}\,.
\end{eqnarray}

The entanglement entropy for each mode is calculated to be
\begin{eqnarray}
S(p,m)&=&-{\rm Tr}\rho_R\log\rho_R\nonumber\\
&=&2\log\left(1+|\gamma_p|^2\right)
-\frac{2|\gamma_p|^2}{1+|\gamma_p|^2}\log|\gamma_p|^2\,.
\label{fermion}
\end{eqnarray}
The final entanglement entropy per unit comoving volume between two causally disconnected regions are obtained by integrating over $p$ and a volume integral over the hyperboloid $H^3$. That is, we use the density of states on the hyperboloid~\cite{Maldacena:2012xp}
\begin{eqnarray}
S(m)=2\pi\int_0^\infty dp\,{\cal D}(p) S(p,m)\,.
\label{integrate}
\end{eqnarray}
The density of states for $H^3$ in the case of the Dirac field is ${\cal D}(p)=(\frac{1}{4}+p^2)/(2\pi^2)$~\cite{Camporesi:1995fb, Bytsenko:1994bc}.
The result is plotted in red line of Figure~\ref{fig2} where $\nu$ is defined in Eq.~(\ref{nu}). We plotted the entanglement entropy for one degree of freedom for comparison with a scalar field, that is, the plot is $1/2$ of Eq.~(\ref{integrate})\footnote{As explained below Eq.~(\ref{cdab}) we focus on two degrees of freedom of a fermion since two other degrees of freedom satisfy the same relations.}.

Now let us compare the result with the entanglement entropy of a scalar field which is computed by~\cite{Maldacena:2012xp} and expressed as
\begin{eqnarray}
S(p,\nu)&=&-\sum_{n=0}^\infty\left(1-|\gamma|^2\right)|\gamma|^{2n}\log\left\{\left(1-|\gamma|^2\right)|\gamma|^{2n}\right\}\nonumber\\
&=&-\log\left(1-|\gamma|^2\right)-\frac{|\gamma|^2}{1-|\gamma|^2}\log|\gamma|^2\,,
\label{scalar}
\end{eqnarray}
where $\gamma$ is given by
\begin{eqnarray}
\gamma = i\frac{\sqrt{2}}{\sqrt{\cosh 2\pi p + \cos 2\pi \nu}
 + \sqrt{\cosh 2\pi p + \cos 2\pi \nu +2 }}\,,
\end{eqnarray}
and a mass parameter is defined by
\begin{eqnarray}
\nu=\sqrt{\frac{9}{4}-\frac{m^2}{H^2}}\,.
\label{nu}
\end{eqnarray}
In the case of a massless scalar field $(\nu=3/2)$, we find $\gamma=e^{-\pi p}$, and then the reduced density matrix is found to be thermal
\begin{eqnarray}
\rho_R=\left(1-e^{-2\pi p}\right)\sum_{n=0}^\infty e^{-2\pi pn}|n\rangle\langle n|\,,
\end{eqnarray}
with temperature $T=H/(2\pi)$.

We integrate over $p$ as in Eq.~(\ref{integrate}). The density of states for $H^3$ in the case of the scalar field is ${\cal D}(p)=p^2/(2\pi^2)$~\cite{Bytsenko:1994bc}. The result of the scalar field is plotted in blue line in Figure~\ref{fig2}.
In Figure~\ref{fig2}, we see that the Dirac field gets more entangled than the scalar field as $m^2/H^2$ becomes small, and the difference is maximal in the massless limit.

\subsection{Fermion seems more entangled than scalar in the massless
limit}
For the massless Dirac field ($\nu^2=9/4$), $\gamma_p=e^{-\pi p}$ which becomes 
$1$ in the limit of $p=0$. Then the entanglement entropy of the Dirac field 
per each degree of freedom, Eq.~(\ref{fermion}), becomes 
$\log 2$.\footnote{We consider two degrees of freedom of a fermion now 
as explained below Eq.~(\ref{cdab})} 
Since the density of states of the Dirac field is finite even in the limit 
of $p=0$, the final entanglement entropy Eq.~(\ref{integrate}) on 
large scales is finite. For a massless scalar field, on the other hand, 
the entanglement entropy Eq.~(\ref{scalar}) becomes logarithmically 
infinite in the limit of $p=0$. But the density of states of the scalar 
field becomes quadratically zero. Then the entanglement entropy summing 
over $p$ gives zero in the limit of $p=0$. 
Thus the contribution of the states from large scales to the entanglement
 entropy becomes large for the massless Dirac field compared to the case of
 the massless scalar field.

However, one should note that the scalar field entanglement entropy shows
a strange behavior as the mass decreases. For $m^2<2H^2$ ($\nu^2>1/4$),
the entanglement entropy once decreases as the mass decreases. It is known
that there exits a supercurvature mode for $m^2<2H^2$, and its contribution
to the long-distance correlation becomes more and more dominant as $m^2/H^2\to0$~\cite{Sasaki:1994yt}.
At the moment we have no clue about how the supercurvature mode affects
the entanglement. It seems possible that the contribution of the
supercurvatue mode, if it could ever be computed, would dominate 
the entanglement entropy of the scalar field in the small mass limit.
This issue needs further studies before we make a firm conclusion.

\section{Summary}

We studied the entanglement entropy of a free massive Dirac field between 
two causally disconnected open charts in de Sitter space. 
For this purpose, we first derived the Bunch-Davies vacuum mode functions 
of the Dirac field in the coordinates that respect the open chart. 
We then gave the Bogoliubov transformation between the Bunch-Davies 
vacuum and the open chart vacua that makes the reduced density matrix diagonal.
 We derived the reduced density matrix in one of the open charts ($R$ region) 
after tracing out the other ($L$ region) and found that the Fermi-Dirac 
distribution is realized in the limit of masslessness.
We then computed the entanglement entropy of the Dirac field by using the 
reduced density matrix. We compared the entanglement entropy of one 
degree of freedom of the Dirac field with that of a scalar field
 calculated by~\cite{Maldacena:2012xp}. 

We found that the entanglement entropy for the Dirac 
field gets more entangled than that for a scalar field as $m^2/H^2$ 
becomes small, and the difference is maximal in the massless limit. 
This is because the contribution of the states from large scales to the 
entanglement entropy becomes large for the massless Dirac field compared 
to the case of the massless scalar field. But there is a caveat.
In the computation of the entanglement entropy of a scalar field,
it is assumed that the supercurvature mode does not contribute.
If this assumption were wrong, the entanglement entropy of the scalar
field might become larger than that of the Dirac fermion in the small mass
limit.
In connection with this issue, we also showed that
there is no supercurvature mode for the Dirac field.

\appendix

\section{Necessary Formulae}
\label{app:a}

\begin{eqnarray}
F\left(\alpha,\beta,\gamma;z\right)&=&\frac{\Gamma(\gamma)\Gamma(\alpha+\beta-\gamma)}{\Gamma(\alpha)\Gamma(\beta)}\left(1-z\right)^{\gamma-\alpha-\beta}F\left(\gamma-\alpha,\gamma-\beta,\gamma-\alpha-\beta+1;1-z\right)\nonumber\\
&&+\frac{\Gamma(\gamma)\Gamma(\gamma-\alpha+\beta)}{\Gamma(\gamma-\alpha)\Gamma(\gamma-\beta)}
F\left(\alpha,\beta,\alpha+\beta-\gamma+1;1-z\right)\,,
\end{eqnarray} 
and
\begin{eqnarray}
F\left(\alpha,\beta,\gamma;z\right)=\left(1-z\right)^{\gamma-\alpha-\beta}
F\left(\gamma-\alpha,\gamma-\beta,\gamma;z\right)\,.
\end{eqnarray}



\section*{Acknowledgments}
We would like to thank Jiro Soda for valuable discussions, suggestions and comments.
SK was supported by IKERBASQUE, the Basque Foundation 
for Science and the Basque Government (IT-559-10),  
and Spanish Ministry MINECO  (FPA2015-64041-C2-1P). 
This work was supported in part by MEXT KAKENHI Nos.~15H05888, 15K21733, 24103001, 24103006, 15H02087 and 26287044.

\end{document}